\RequirePackage{fix-cm}
\documentclass[10pt]{article}
\usepackage{graphicx}
\usepackage{multirow}

\usepackage[margin=2cm]{geometry}
\usepackage{hyperref}

\begin{document}

\title{A systematic review and meta-analysis of interaction models between transportation networks and territories}






\author{Juste Raimbault$^{1,2,3,\ast}$\medskip\\
$^{1}$ Center for Advanced Spatial Analysis, University College London\\
$^{2}$ UPS CNRS 3611 ISC-PIF\\
$^{3}$ UMR CNRS 8504 G{\'e}ographie-cit{\'e}s\medskip\\
$^{\ast}$ \texttt{juste.raimbault@polytechnique.edu}
}

\date{}


\maketitle



\begin{abstract}
Modeling and simulation in urban and regional studies has always given a significant place to models relating the dynamics of territories with transportation networks. These include for example Land-use Transport Interaction models, but this question has been investigated from different viewpoints and disciplines. We propose in this paper a systematic review to construct a corpus of such models, followed by a meta-analysis of model characteristics. A statistical analysis provides links between temporal and spatial scale of models, their level of interdisciplinarity, and the paper year, with disciplines, type of model and methodology. We unveil in particular strong disciplinary discrepancies in the type of approach taken. This study provides a basis for novel and interdisciplinary approaches to modeling interactions between transportation networks and territories.
\end{abstract}

\section{Introduction}

Within urban and territorial systems, infrastructure networks constitute a backbone for the dynamics of agents and components. Transportation networks have thus a strongly shaping role in urban dynamics, at small time scales by sustaining daily mobility \cite{zhong2015measuring}. At longer time scales, transport networks determine accessibility patterns which in turn influence residential and employment relocations \cite{wegener2004land}. Computational models have been introduced for a long time in the literature to understand, anticipate and manage such interactions between transportation networks and territories.

Such models stem from diverse disciplines and answer different research questions, focusing on various processes and dimensions of interactions. In economics for example, the impact of network on urban form is studied by accounting for the location of economic sectors \cite{baum2017roads}. The role of networks in urban growth is also investigated from the economic viewpoint \cite{baum2020does}. In transportation or urban planning, the focus is made on changes in accessibility landscapes \cite{santos2010interurban}. Operational urban models such as Land-use Transport Interaction models \cite{wegener2004land} are also closely related to the discipline of planning. At other scales, geography focuses on the dynamics of urban systems and the role of networks within \cite{pumain2017urban}, but also on the relation between road network topology and urban form \cite{raimbault2019urban}. The field of transportation science, including transportation geography, ranges from accessibility studies \cite{geurs2004accessibility} to transportation and traffic models \cite{mcnally2007four}.


Altogether, this wide variety of approaches to the same conceptual object makes it difficult to overview them synthetically, to relate them and the respective disciplines, and to potentially investigate for new and interdisciplinary modeling approaches. \cite{raimbault2018caracterisation} provides a literature mapping in terms of citation network and semantic network, giving already a first insight into this issue. However, the characteristics of models themselves and the research questions answered are not studied in details. We propose in this paper to dig deeper into model properties, in relation with the disciplinary context, in a systematic way. Our contribution is thus twofold: (i) through a systematic review, we build an interdisciplinary corpus of 145 papers studying interaction models between transportation networks and territories; (ii) after extracting models properties, we proceed to a meta-analysis by estimating statistical models searching for significant covariates of these properties.

The rest of this paper is organised as follows: we first describe the methods used both for the systematic review and the meta-analysis; we then give descriptive results, both qualitative observations and quantitative measures, stemming from the systematic review; we proceed to the meta-analysis with statistical models; we finally discuss the implications of our results in terms of building new interaction models.

\section{Methods}

\subsection{Systematic review}

Systematic literature reviews mostly take place in fields where a very targeted request, even by article title, will yield a significant number of studies studying quite the same question: typically in therapeutic evaluation, where standardised studies of a same molecule differ only by the size of samples and statistical modalities (control group, placebo, level of blinding). The PRISMA reporting guidelines for systematic reviews enter this frame \cite{moher2009preferred}. In this case corpus construction is straightforward first thanks to the existence of specialised bibliographic databases allowing very precise requests, and furthermore thanks to the possibility to proceed to additional statistical analyses to confront the different studies (for example network meta-analysis, see~\cite{rucker2012network}). In our case, the exercise is less direct for the reasons exposed above and others: objects are hybrid, problematics are diverse, and disciplines are numerous. The different points we will raise in the following will often have as much thematic value as methodological value, suggesting crucial points for the realisation of such an hybrid systematic review.

We propose an hybrid methodology coupling the two methodologies developed in \cite{raimbault2017models} and \cite{raimbault2019exploration}, with a more classical procedure of systematic review. We aim both at a representativity of all the disciplines we discovered, but also a limited noise in the references taken into account for the meta-analysis of model characteristics. Therefore, we combine the corpus obtained in the literature mapping through citation network analysis by \cite{raimbault2018caracterisation} and a corpus constructed through keywords requests, in a way similar to \cite{tahamtan2018core}. The protocol is thus the following:

\begin{enumerate}
	\item Starting from the citation corpus constructed by \cite{raimbault2018caracterisation}, we isolate a number of relevant keywords, by selecting the 5\% of links having the strongest weight (arbitrary threshold), and among the corresponding nodes the ones having a degree larger than the quantile at 0.8 of their respective semantic class. The first filtration allows to focus on the ``core'' of observed disciplines, and the second to not introduce size bias without loosing the global structure, classes being relatively balanced. A manual screening allows to remove keywords that are obviously not relevant (remote sensing, tourism, social networks, \ldots), what leads to a corpus of $K=115$ keywords ($K$ is endogenous here).
	\item For each keyword, we automatically do a catalog request (google scholar) while adding \texttt{model*} to it, of a fixed number $n=20$ of references. The supplementary term is necessary to obtain relevant references, after testing on samples.
	\item The potential corpus composed of obtained references, with references composing the citation network, is manually screened (review of titles) to ensure a relevance regarding the state of the art reviewed extensively by \cite{raimbault2018caracterisation}, yielding the preliminary corpus of size $N_p = 297$.
	\item This corpus is then inspected for abstracts and full texts if necessary. We select articles elaborating a modeling approach, ruling out conceptual models. References are classified and characterised according to criteria described below. We finally obtain a final corpus of size $N_f = 145$, on which quantitative analyses are possible.
\end{enumerate}

\begin{figure}
\begin{center}
\frame{\includegraphics[width=0.8\linewidth]{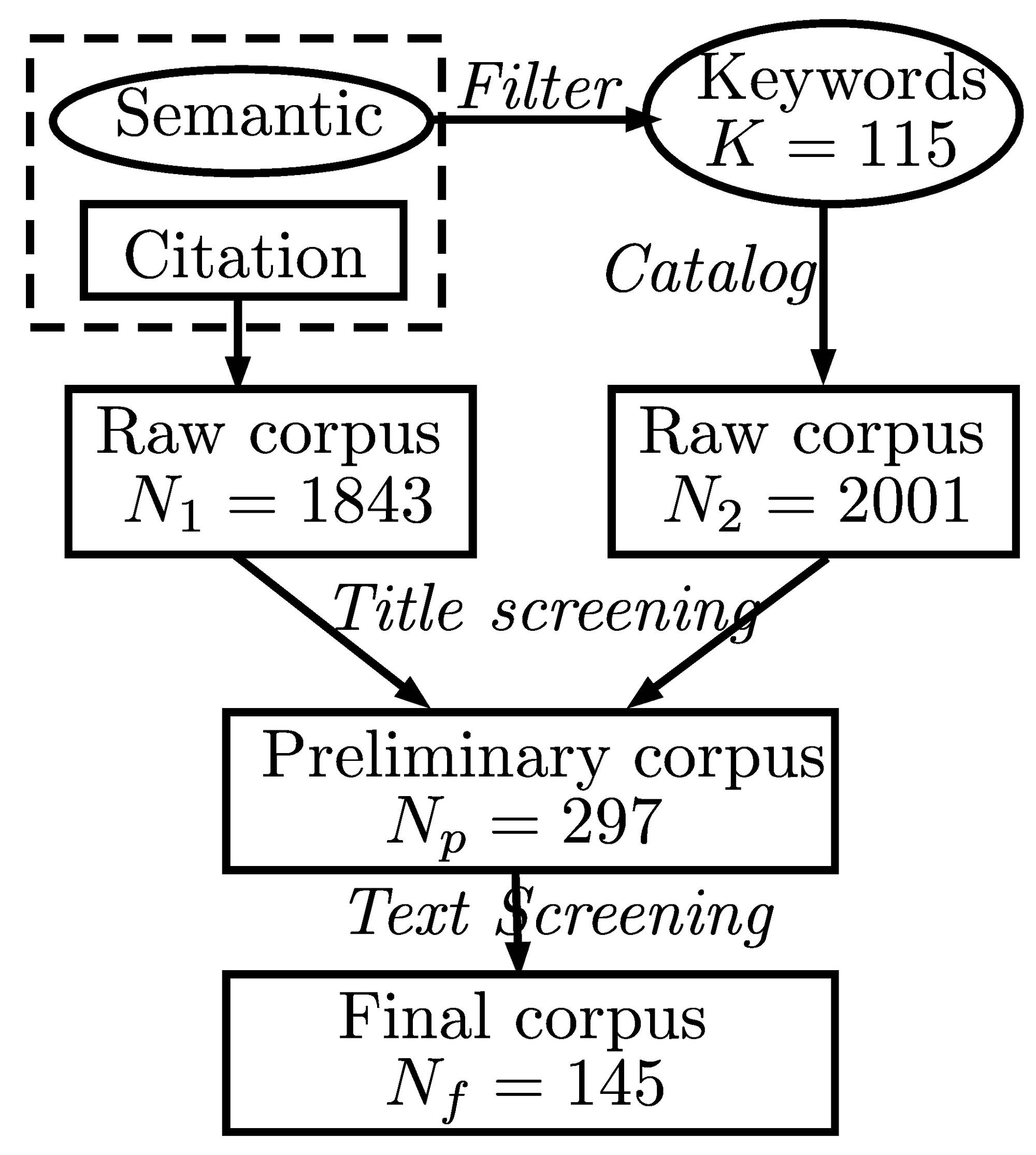}}
\end{center}
\caption{\textbf{Methodology of the systematic review.} Rectangles correspond to corpuses of references, ellipses to corpuses of keywords, and dashed lines to initial corpuses. At each stage the size of the corpus is given.\label{fig:modelography:systematicreview}}
\end{figure}

The method is summarised in Fig.~\ref{fig:modelography:systematicreview}, with parameter values and the size of the successive corpuses.

For the choice of initial keywords for the indirect construction (through semantic request), a possible alternative could be to extract the relevant keywords for each sub-community of the citation network, and then select the most relevant for each domain. We make the choice of extracting them on the complete corpus, and then to collect them by sub-community thereafter. For a small corpus, the second option is more suitable, since the notion of relevance is less importance than for very large corpuses, in which some relevant words may be drowned and less relevant to emerge in a spurious way. In other words, the keyword selection method appears to be more robust on smaller corpuses, as suggest the comparison of this application with the one done on the Cybergeo journal by~\cite{raimbault2019exploration} and the one done on the patent corpus by \cite{bergeaud2017classifying}.

The methods used do not allow to avoid some ``noise'', i.e. retrieving articles that are not relevant to the subject, even with a very low tolerance threshold. We obtained for example articles from unrelated disciplines such as gender studies, biology, criminology, urban geology.
 This confirms that the manual filtering stage is essential.

This noise can for example be due to:
\begin{itemize}
	\item effective citations for diverse reasons, but having only a low relevance in the citing article;
	\item noise intrinsic to the keyword search;
	\item catalog classification errors.
\end{itemize}

\subsection{Meta-analysis}

The second part of our study is devoted to a mixed analysis based on the corpus constructed through the systematic review. 
 It aims at extracting and to precisely decompose ontologies, scales and processes, and then to study possible links between these characteristics of the models and the context in which they have been introduced. It is thus in a way the meta-analysis, that we can also designate as ``modelography'' following \cite{schmitt2013modelographie}. It is indeed not a meta-analysis strictly speaking since we do not combine similar analysis to extrapolate potential results from larger samples. Our approach is close to the one of \cite{cottineau2017metazipf} which gathers references having quantitatively studied Zipf's law for cities, and then links the characteristics of studies to the methods used and the assumptions formulated.

The first part consists in the extraction of the characteristics of models. An automation this work would consist in a research project in itself, as we develop in discussion below, but we are convinced of the relevance to refine such techniques 
 in the context of developing integrated disciplines. We focus here on a manual extraction that will aim at being more precise that an approximatively convincing data mining attempt.
 
\begin{figure}
\begin{center}
\includegraphics[width=\linewidth]{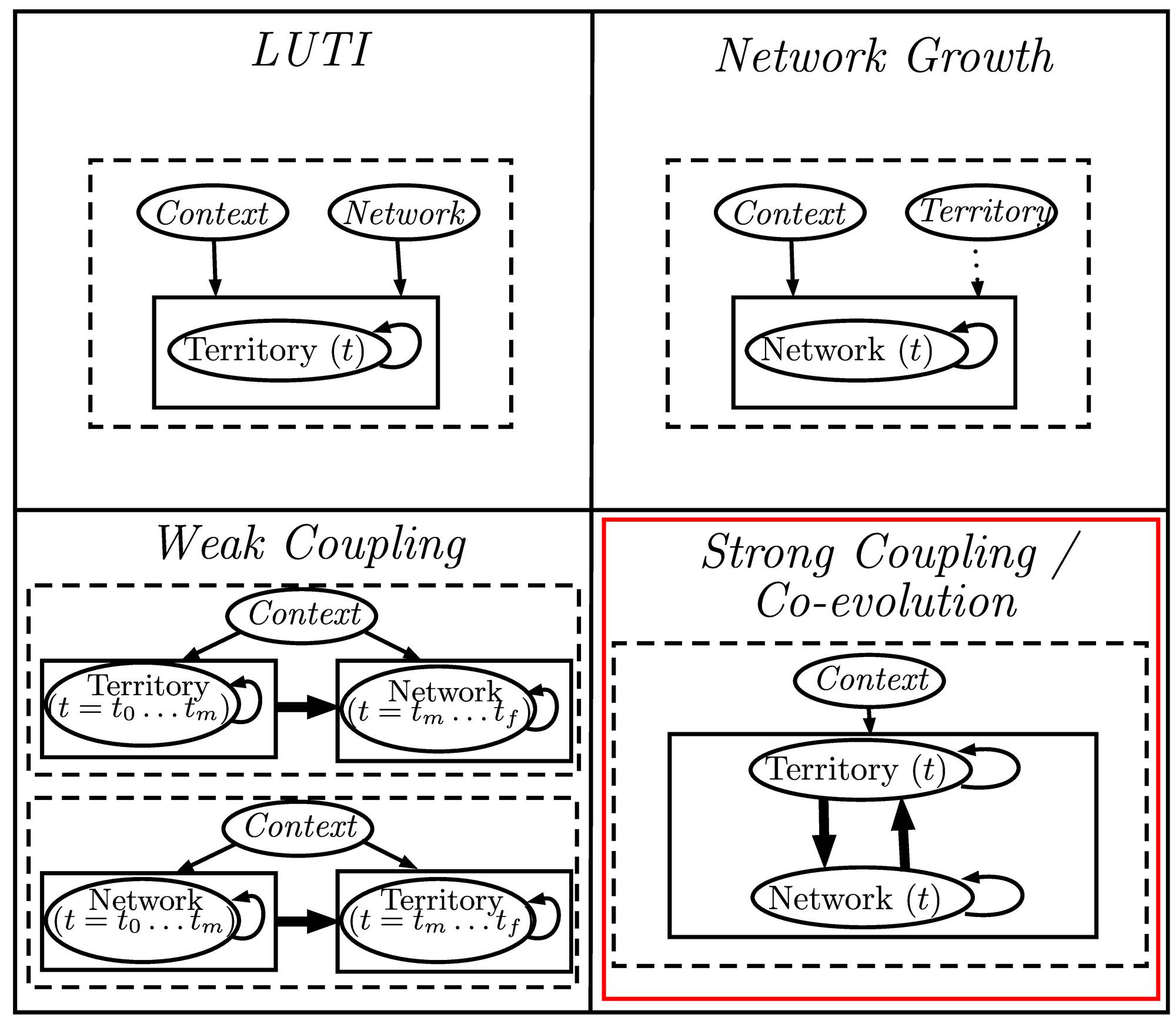}
\end{center}
\caption{\textbf{Schematic representation of the distinction between different types of models coupling networks and territories.} This typology is based on the one of \cite{raimbault2018caracterisation}, by distinguishing approaches in which territory or network are given as a context (Luti and network growth) from a sequential coupling between a model for each. Ontologies are represented as ellipses, submodels by full boxes, models by dashed boxes, couplings by arrows. We highlight in red the co-evolution approach which remains rare in the literature and has significant development potential.\label{fig:modelography:coevolution}}
\end{figure}

We extract from models the following characteristics:

\begin{itemize}
	\item what is the strength of coupling between territorial ontologies and the ones of the network, in other words is it a co-evolution model in the sense of \cite{raimbault2018caracterisation} or a weak coupling. To the best of our knowledge there does not exist generic approaches to model coupling that would be not linked to a particular formalism. We will take the approach given in introduction of \cite{raimbault2018caracterisation}, by distinguishing here a weak coupling as a sequential coupling (outputs of the first model become inputs of the second) from a dynamical strong coupling where the evolution is interdependent at each time step (either by a reciprocal determination of by a common ontology). We will therefore classify into categories following the representation of figure~\ref{fig:modelography:coevolution}: \texttt{\{territory ; network ; weak ; coevolution\}}, which results from the analysis of literature in~\cite{raimbault2018caracterisation};
	\item maximal time scale;
	\item maximal spatial scale;
	\item domain ``a priori'', determined by the origin of authors and the domain of the journal;
	\item methodology used (statistical models, system of equations, multi-agent, cellular automaton, operational research, simulation, etc.);
	\item case study (city, metropolitan area, region or country) when relevant.
\end{itemize}

We also collect in an indicative way, but without objective for objectivity nor exhaustivity, the ``subject'' of the study (i.e. the main thematic question) ans also the ``processes'' included in the model. An exact extraction of processes remains hypothetical, on the one hand because it is conditioned to a rigorous definition and taking into account different levels of abstraction, of complexity, or scale, on the other hand as it depends on technical means out of reach of this modest study. We will comment these in an indicative way without including them in systematic studies.

We also gather scale, range and in a sense resolution to not make the extraction more complicated. Even if it would be relevant to differentiate when an element does not exist for a model (NA) to when it is badly defined by the author, this task seem to be sensitive to subjectivity and we merge the two modalities.

Based on the citation network mapping done in \cite{raimbault2018caracterisation}, we add to the previous characteristics the following variables:

\begin{itemize}
	\item citation domain (when available, i.e. for references initially present in the citation network, what corresponds to 55\% of references);
	\item semantic domain, defined by the domain for which the document has the highest probability;
	\item index of interdisciplinarity.
\end{itemize}

Semantic domains and the interdisciplinarity measure have been recomputed for this corpus through the collection of keywords, and then keyword extraction following the method described in~\cite{raimbault2019exploration}, with $K_W=1000$, $\theta_w=15$ and $k_{max}=500$. We obtain more targeted communities which are relatively representative of thematics and methods: Transit-oriented development (\texttt{tod}), Hedonic models (\texttt{hedonic}), Infrastructure planning (\texttt{infra planning}), High-speed rail (\texttt{hsr}), Networks (\texttt{networks}), Complex networks (\texttt{complex networks}), Bus rapid transit (\texttt{brt}).

An ``appropriate choice'' of characteristics to classify models is similar to the issue of choosing features in machine learning: in the case of supervised learning, i.e. when we aim at obtaining a good prediction of classes fixed a priori (or a good modularity of the obtained classification relatively to the fixed classification), we can select features optimising this prediction. We will therein discriminate models that are known and judged different. If we want to extract an endogenous structure without any a priori (unsupervised classification), the issue is different. We will therefore test in a second time a regression technique which allows to avoid overfitting and to select features (random forests).

\section{Results}

To construct the corpus, we used the open source tools developed by \cite{raimbault2019exploration} and available at \url{https://github.com/JusteRaimbault/BiblioData}. Corpus data and processing code for the meta-analysis are openly available at \url{https://github.com/JusteRaimbault/CityNetwork/tree/master/Models/QuantEpistemo/HyperNetwork}. The R software was used for preparing the data, network analysis and statistical models.

\subsection{Corpus construction}

During the manual classification achieved when screening abstracts, the following points are of interest in terms of qualitative methodological results: (i) the ``a priori'' disciplines are judged based on the journal in which the article was published - in particular, we operate the following particular choices (for other journals such as physics journals there is no ambiguity): Journal of Transport Geography, Environment and Planning B: Geography; Journal of Transport and Land-use, Transportation Research: Transportation; (ii) geography in our sense includes urbanism and urban studies is these are not too close from planning (urban sustainability for example).

The systematic review procedure first of all allows revealing several methodological points, which knowledge can be an asset to proceed to similar hybrid systematic reviews:

\begin{itemize}
	\item Catalog bias seem to be inevitable. We rely on the assumption that the use of google scholar allows a uniform sampling regarding catalog errors or bias. The future development of open tools for cataloging and mapping, allowing contributed efforts for a more precise knowledge of extended fields and of their interfaces, is a crucial issue for the reliability of such methods as illustrated by~\cite{raimbault2019empowering}.
	\item The availability of full texts is an issue, in particular for such a broad review, given the multiplicity of editors. The existence of tools to emancipate science such as Sci-hub (\url{http://sci-hub.cc/}) allows effectively accessing full texts. Echoing the recent debate on the negotiations with publishers regarding the exclusivity of full texts mining, it appears to be more and more salient that a reflexive open science is totally orthogonal to the current model of publishing. We also hope for a rapid evolution of practices on this point.
	\item Journals, and indeed publishers, seem to differently influence the referencing, potentially increasing the bias during requests. Grey literature and preprints are taken into account in different ways depending on the domains.
	\item Manual screening of large corpora allows to not miss ``crucial papers'' that could have been omitted before~\cite{lissack2013subliminal}. The issue of the extent to which we can expect to be informed in the most exhaustive way possible of recent discoveries linked to the subject studied is very likely to evolve given the increase of the total amount of literature produced and the separation of fields, among which some are always more refined~\cite{bastian2010seventy}. Following the previous points, we can propose that tools helping systematic analysis will allow to keep this objective as reasonable.
	\item Results of the automatised review are significantly different from the domains highlighted in a classical literature review: some conceptual associations, in particular the inclusion of network growth models, are not natural and do not exist much in the scientific landscape as we previously showed.
\end{itemize}

Furthermore, the operation of constructing the corpus already allows to draw thematic observations that are interesting in themselves:

\begin{itemize}
	\item The articles selected imply a clarification of what is meant by ``model''. In the sense of \cite{raimbault2017applied}, a very broad definition of model which applies to any scientific perspectives can be given. Our selection here does not retain conceptual models for example, our choice criteria being that the model must include a numerical or simulation aspect.
	\item A certain number of references consist in reviews, what is equivalent to a group of model with similar characteristics. 
	We could make the method more complicated by transcribing each review or meta-analysis, or by weighting the records of corresponding characteristics by the corresponding number of articles. We make the choice to ignore these reviews, what remains consistent in a thematic way still with the assumption of uniform sampling.
	\item A first clarification of the thematic frame is achieved, since we do not select studies uniquely linked to traffic and mobility (this choice being also linked to the results obtained by~\cite{raimbault2017investigating}), to pure urban design, to pedestrian flows models, to logistics, to ecology, to technical aspects of transportation, to give a few examples, even if these subjects can from a specific viewpoint be considered as linked to interactions between networks and territories.
	\item Similarly, neighbour fields such as tourism, social aspects of the access to transportation, anthropology, were not taken into account.
	\item We observe a high frequency of studies linked to High Speed Rail (HSR), recalling the necessary association of political aspects of planning and of research directions in transportation.
\end{itemize}

\subsection{Meta-analysis}

\subsubsection{Descriptive statistics}

\paragraph{Processes and case study}

Regarding the existence of a case study and its localisation, 26\% of studies do not have any, corresponding to an abstract model or toy model (close to all studies in physics fall within this category). Then, they are spread across the world, with however an overrepresentation of Netherlands with 6.9\%. Processes included are too much varied (in fact as much as ontologies of concerned disciplines) to be the object of a typology, but we will observe the dominance of the notion of accessibility (65\% of studies), and then very different processes ranging from real estate market processes for hedonic studies, to relocations of actives and employments in the case of Luti, or to network infrastructure investments. We observe abstract geometric processes of network growth, corresponding to works in physics. Network maintenance appears in one study, as political history does. Abstract processes of agglomeration and dispersion are also at the core of several studies. Interactions between cities are a minority, the systems of cities approaches being drown in accessibility studies. Issues of governance and regulation also emerge, more in the case of infrastructure planning and of TOD approaches evaluation models, but remain a minority. We will stay with the fact that each domain and then each study introduces its own processes with are quasi-specific to each case.

\paragraph{Corpus characteristics}

The domains ``a priori'' (i.e. judged, or more precisely inferred from journal or institution to which authors belong), are relatively balanced for the main disciplines already identified: 17.9\% Transportation, 20.0\% Planning, 30.3\% Economics, 19.3\% Geography, 8.3\% Physics , the rest in minority being shared between environmental science, computer science, engineering and biology. Regarding the share of significant semantic domains, TOD dominates with 27.6\% of documents, followed by networks (20.7\%), hedonic models (11.0\%), infrastructure planning (5.5\%) and HSR (2.8\%). Contingence tables show that Planning does almost only TOD, physics only networks, geography is equally shared between networks and TOD (the second corresponding to articles of the type ``urban project management'', that have been classified in geography as published in geography journals) and also a smaller part in HSR, and finally economics is the most diverse between hedonic models, planning, networks and TOD. This interdisciplinarity however appears only for classes extracted for the higher probability, since average interdisciplinarity indices by discipline have comparable values (from 0.62 to 0.65), except physics which is significantly lower at 0.56 what confirms its status of ``newcomer'' with a weaker thematic depth.

\paragraph{Models studied}

It is interesting for our problematic to answer the question ``who does what ?'', i.e. which type of models are used by the different disciplines. We give in Table~\ref{tab:modelography:what} the contingency table of the type of model as a function of a priori disciplines, of the citation class and of the semantic class. We observe that strongly coupled approaches, the closest of what is considered as co-evolution models, are mainly contained in the vocabulary of networks, what is confirmed by their positioning in terms of citations, but that the disciplines concerned are varied. The majority of studies focus on the territory only, the strongest unbalance being for studies semantically linked to TOD and hedonic models. Physics is still limited as focusing exclusively on networks.

\begin{table}
\caption{\textbf{Types of models studied according to the different classifications.} Contingence tables of the discrete variable giving the type of model (network, territory or strong coupling), for the a priori classification, the semantic classification and the citation classification.\label{tab:modelography:what}}\medskip
\begin{center}
\begin{tabular}{|p{2.4cm}|p{2.4cm}p{2.4cm}p{2.4cm}p{2.4cm}p{2.4cm}|}
\hline
Discipline  &  economics & geography & physics & planning & transportation\\\hline
network     &     5      &      3    &   12    &    1     &         4  \\
strong      &     4      &     3     &   0     &   0      &        2  \\
territory   &    35      &    22     &   0     &    28    &         20 \\\hline  
\end{tabular}

\medskip

\begin{tabular}{|p{2.4cm}|p{2.4cm}p{2.4cm}p{2.4cm}p{2.4cm}p{2.4cm}|}
\hline
Semantic  &  hedonic & hsr & infra planning & networks & tod\\\hline
network   &       1  & 0   &          0     &  14      & 2 \\
strong    &       0  &  0  &            0   &     5    & 1  \\
territory &      15  &  4  &            8   &    11    &  37 \\ \hline
\end{tabular}

\medskip

\begin{tabular}{|p{2cm}|p{2cm}p{2cm}p{2cm}p{2cm}p{2cm}p{2cm}|}
\hline
Citation  &  accessibility & geography & infra planning & LUTI & networks & TOD \\\hline
network   &            0   &     0     &         0      &   0  &     24   &  0 \\
strong    &            0   &      0    &          0     &   2  &      5   &  0 \\
territory &           13   &      1    &          6     &  18  &      2   &  3 \\\hline
\end{tabular}
\end{center}
\end{table}

\paragraph{Studied scales}

To then answer the question of the how, we can have a look at temporal and spatial typical scales of models. Planning and transportation are concentrated at small spatial scales, metropolitan or local, economics also with a strong representation of the local through hedonic studies, and a spatial range a bit larger with the existence of studies at the regional level and a few at the scale of the country (panel studies generally). Again, physics remains limited with all its contributions at a fixed scale, the metropolitan scale (which is not necessarily clear nor well specified in articles in fact since these are toy models which thematic boundaries may be very fuzzy). Geography is relatively well balanced, from the metropolitan to the continental scale. The scheme for temporal scales is globally the same. The methods used are strongly correlated to the discipline: a $\chi^2$ test gives a statistic of 169, highly significant with $p=0.04$. Similarly, spatial scale also is but in a less strong manner ($\chi^2 = 50, p = 0.08$).

\subsection{Statistical analysis}

\subsubsection{Linear models with model selection}

We now study the influence of diverse factors on characteristics of models through simple linear regressions. In a multi-modeling approach, we propose to test all the possible models to explain each of the variables from the others. The number of observations for which all the variables have a value is very low, we need to take into account the number of observations used to fit each model. Furthermore, model performances can be characterised by complementary objectives. Following~\cite{igel2005multi}, we apply a multi-objective optimization, to simultaneously maximise the explained variance (adjusted R$^2$ in our case) and the information captured (corrected Akaike information criterion AICc - AIC is a measure of the information gain between two models, and allows to avoid abusive overfitting through a too large number of parameters; AICc is a version taking into account the size of the sample, the measure varying significantly for the small samples). It is realised conditionally to the fact of having the number of observations $N>50$ (fixed threshold regarding the distribution of $N$ on all models). The optimization procedure is detailed in Supplementary Material for each variable. Time scale and interdisciplinarity exhibit compromises difficult choose from, and we adjust the two candidates. Other variables exhibit dominating solutions and we adjust only a single model.

Complete regression results are given in Table~\ref{tab:quantepistemo:regressions}. Temporal and spatial scales, together with year, are the variables the best explained in the sense of the variance. Time scale is very significantly influenced by the type of model: territory which decreases it, or strong coupling which increases it. The fact to be in physics also significantly influences, and broadens the time range of models. On the contrary, engineering approaches (often optimal design of a transportation network) correspond to a short time span.

For the spatial scale, the fact to be in geography has a strong influence on the spatial range of models: indeed, regional studies and at the scale of the system of cities are indeed the prerogative of geography. The belonging to the field of transportation also increases slightly the spatial range (see significance in the complete regressions in Supplementary Material). No other variable has a significant influence.

The level of interdisciplinarity is well explained by the year, which influences it in a negative way, what confirms an increase in scientific specialisations in time. Econometric studies of hedonic models appears to be very specialised. Finally, publication year is significantly and positively explained by the territory type and by the fact to be in transportation, what would correspond to a recent resurgence of a particular profile of studies. A study of the corpus suggests that this would be studies on high speed, which would appear as a recent scientific fashion.

\begin{table}
\caption{\textbf{Explanation of models characteristics.} Results of the Ordinary Least Squares (OLS) estimation of selected linear models, for each variable to be explained: temporal scale (TEMPSCALE), spatial scale (SPATSCALE), interdisciplinarity index (INTERDISC), publication year (YEAR).\label{tab:quantepistemo:regressions}}
\begin{tabular}{lcccccc} 
\footnotesize
\\[-1.8ex]\hline 
\hline \\[-1.8ex] 
 & \multicolumn{6}{c}{\textit{Explained variable:}} \\ 
\cline{2-7} 
 & \multicolumn{2}{c}{TEMPSCALE} & SPATSCALE & \multicolumn{2}{c}{INTERDISC} & YEAR \\ 
 & (1) & (2) & (3) & (4) & (5) & (6)\\ 
\hline \\[-1.8ex] 
 YEAR & 0.674 &  &  & $-$0.004$^{*}$ & $-$0.002$^{*}$ &  \\ 
  TYPEstrong &  & 100.271$^{***}$ &  &  & $-$0.026 &  \\ 
  TYPEterritory & $-$38.933$^{***}$ & $-$14.988 &  &  & 0.044 & 10.898$^{***}$ \\ 
  TEMPSCALE &  &  & $-$5.179 & $-$0.0003 &  & 0.035 \\ 
  FMETHODeq &  &  &  &  &  & $-$6.224 \\ 
  FMETHODmap &  &  &  &  &  & 4.747 \\ 
  FMETHODro &  &  &  &  &  & 6.128 \\ 
  FMETHODsem &  &  &  &  &  & 1.009 \\ 
  FMETHODsim &  &  &  &  &  & 5.153 \\ 
  FMETHODstat &  &  &  &  &  & $-$0.357 \\ 
  DISCIPLINEengineering & $-$52.107$^{*}$ & $-$9.609 & $-$154.461 & 0.144 &  & 13.486 \\ 
  DISCIPLINEenvironment & 17.110 & 17.886 & $-$5.878 & 0.092 &  & $-$3.668 \\ 
  DISCIPLINEgeography & 3.640 & 9.126 & 1,445.457$^{***}$ & 0.036 &  & 1.121 \\ 
  DISCIPLINEphysics & 46.879$^{*}$ & 77.897$^{***}$ & 292.559 & $-$0.103 &  & 3.392 \\ 
  DISCIPLINEplanning & 1.304 & 4.553 & $-$143.554 & $-$0.047 &  & $-$2.850 \\ 
  DISCIPLINEtransportation & $-$14.718 & 8.753 & 568.329 & 0.062 &  & 5.503$^{*}$ \\ 
  INTERDISC & 2.357 &  &  &  &  & $-$12.876 \\ 
  SEMCOMcomplex networks &  &  &  &  & $-$0.217 &  \\ 
  SEMCOMhedonic &  &  &  & $-$0.179 & $-$0.184$^{*}$ & $-$5.769 \\ 
  SEMCOMhsr &  &  &  & $-$0.100 & $-$0.122 & 6.135 \\ 
  SEMCOMinfra planning &  &  &  & $-$0.032 & $-$0.096 & $-$4.123 \\ 
  SEMCOMnetworks &  &  &  & $-$0.038 & $-$0.107 & 4.711 \\ 
  SEMCOMtod &  &  &  & $-$0.105 & $-$0.152 & $-$1.653 \\ 
  Constant & $-$1,305.126 & 22.103$^{*}$ & 235.357 & 8.962$^{**}$ & 5.531$^{**}$ & 2,004.945$^{***}$ \\ 
 \hline \\[-1.8ex] 
Observations & 64 & 94 & 94 & 64 & 98 & 64 \\ 
R$^{2}$ & 0.385 & 0.393 & 0.100 & 0.314 & 0.155 & 0.510 \\ 
Adj. R$^{2}$ & 0.282 & 0.336 & 0.027 & 0.136 & 0.068 & 0.281 \\ 
\hline 
\hline \\[-1.8ex] 
\textit{Note:}  & \multicolumn{6}{l}{$^{*}$p$<$0.1; $^{**}$p$<$0.05; $^{***}$p$<$0.01} \\ 
\end{tabular}
\end{table} 

\subsubsection{Random Forest regressions}

We conclude this study by regressions and classification with random forests, which are a very flexible method allowing to unveil a structure from a dataset~\cite{liaw2002classification}. To complement the previous analysis, we propose to use it to determine the relative importances of variables for different aspects. We use each time forests of size 100000, a node size of 1 and a number of sampled variables in $\sqrt{p}$ for the classification and $p/3$ for the regression when $p$ is the total number of variables. To classify the type of models, we compare the effects of discipline, of the semantic class and of the citation class. The latest is the most important with a relative measure of 45\%, whereas the discipline accounts for 31\% and the semantic of 23\%. This way, the disciplinary compartmentalization is found again, whereas the semantic and this partly ontologies, is the most open. This encourages us in our aim at getting out of this compartmentalization. When we apply a forest regression on interdisciplinarity, still with these three variables, we obtain that they explain 7.6\% of the total variance, what is relatively low, witnessing a semantic disparity on the whole corpus independently of the different classifications. In this case, the most important variable is the discipline (39\%) followed by the semantic (31\%) and citation (29\%), what confirms that the journal targeted strongly conditions the behavior in the language used. This alerts on the risk of a decrease in semantic wealth when targeting a particular public. This way, we have unveiled certain structures and regularities of models related to the question of interactions between transportation networks and territories, which implications could prove useful during the construction of novel modeling approaches.

\section{Discussion}

\subsection{Developments}

A possible development could consist in the construction of an automatised approach to this meta-analysis, from the point of view of modular modeling, combined to a classification of the aim and the scale. Modular modeling consists in the integration of heterogeneous processes and the implementation of these processes in the aim of extracting mechanisms giving the highest proximity to empirical stylised facts or to data~\cite{cottineau2015growing}. The idea would be to be able to automatically extract the modular structure of existing models, starting from full texts as proposed in~\cite{raimbault2019empowering}, in order to classify these bricks in an endogenous way and to identify potential couplings for new models.

Altogether, meta-analysis protocols in the case of multiple disciplines and multiple models remain to be elaborated. Similarly, being able to extrapolate synthetic results from heterogenous model results remains an open issue.

\subsection{Lessons for new modeling approaches}

We can summarize the main points obtained from this meta-analysis that could influence choices made towards novel modeling approaches. First of all, the interdisciplinary presence of approaches realising a strong coupling confirms our need to build bridges and to couple approaches, and also retrospectively confirms the conclusions of~\cite{raimbault2018caracterisation} on the consequence of discipline compartmentalization in terms of the models formulated. Secondly, the importance of the vocabulary of networks in a large part of models will lead us to confirm this anchorage. The specificity of TOD and accessibility approaches, relatively close to the LUTI models, will be of secondary importance for us. The restricted span of works from physics, confirmed by the majority of criteria studied, suggests to remain cautious of these works and the absence of thematic meaning in the models. The wealth of temporal and spatial scales covered by geographical and economical models confirms the importance of varying these in our models, ideally to reach multi-scale models. Finally, the relative importance of classification variables on the type of model also suggest the direction of interdisciplinary bridges to cross ontologies.

\subsection{Synthesis of modelled processes}

We finally propose to synthesise processes taken into account by models encountered during the meta-analysis, in order to proceed to a similar effort than the one concluding the thematic approach of \cite{raimbault2018caracterisation}. We can neither have an exhaustive view (as already mentioned in the methodology above) nor render with a high precision each model in the details, since almost each is unique in its ontology. The exercise of the synthesis allows then to take a step back from this limits and take a certain height, and have thus an overview on \emph{modeled processes} (keeping in mind selection choices, which lead for example to not have mobility processes within this synthesis).

\begin{table}
\caption{\textbf{Synthesis of modeled processes.} These are classed by scale, type of model and discipline. \label{tab:modelography:processes}}
\begin{tabular}{|l|p{5cm}|p{5cm}|p{5cm}|}
\hline
 & Networks $\rightarrow$ Territories & Territories $\rightarrow$ Networks & Networks $\leftrightarrow$ Territories\\ \hline
\multirow{2}{*}{Micro} &
\textbf{Economics: } real estate market, relocalization, employment market & NA & \textbf{Computer Science : } spontaneous growth \\\cline{2-2}
& \textbf{Planning: } regulations, development & & \\\hline
& \textbf{Economics: } real estate market, transportation costs, amenities & \textbf{Economics: } network growth, offer and demand & \textbf{Economics: } investments, relocalizations, offer and demand, network planning\\\cline{2-4}
\multirow{2}{*}{Meso}& \textbf{Geography: } land-use, centrality, urban sprawl, network effects & \textbf{Transportation: } investments, level of governance & \textbf{Geography: } land-use, network growth, population diffusion \\\cline{2-3}
& \textbf{Planning/transportation: } accessibility, land-use, relocalization, real estate market & \textbf{Physics: } topological correlations, hierarchy, congestion, local optimization, network maintenance & \\\hline
& \textbf{Economics: } economic growth, market, land-use, agglomeration, sprawl, competition & \textbf{Economics: } interactions between cities, investments & \textbf{Economics: } offer and demand \\ \cline{2-4}
\multirow{2}{*}{Macro} & \textbf{Geography: } accessibility, interaction between cities, relocalization, political history & \textbf{Geography: } interactions between cities, potential breakdown & \textbf{Transportation: } network coverage \\\cline{2-3}
& \textbf{Transportation: } accessibility, real estate market & \textbf{Transportation: } network planing & \\\hline
\end{tabular}
\end{table}

The table~\ref{tab:modelography:processes} proposes this synthesis from the 145 articles obtained from the modelography and for which a classification of the type was possible, i.e. that there existed a model entering the typology recalled above. Being fully exhaustive would be similar to an interdisciplinary meta-modeling approach which is far out of the reach of our work
, and the list given here remain indicative.

We find again the correspondences between disciplines, scales and types of models obtained in statistical results above. We still observe the principal lessons, echoing the synthesis table obtained in \cite{raimbault2018caracterisation}:
\begin{enumerate}
	\item The dichotomy of ontologies and processes taken into account between scales and between types is even more explicit here in models than in processes in themselves. Since this study was more detailed, it also appears stronger, since a greater precision allows exhibiting abstract categories. We postulate that there indeed exists different processes at the different scales, and we suggest that multi-scale models or companion models at different scales should be investigated \cite{rozenblat2018conclusion}.
	\item The compartmentalization of disciplines shown by \cite{raimbault2017models} can be found in a qualitative way in this synthesis: it is clear that they originally diverge in their different founding epistemologies. New approaches should aim at integrating paradigms from different disciplines, while taking into account the limits imposed by modeling principles (for example, the parsimony of models necessarily limits the integration of heterogenous ontologies).
	\item An important gap between this synthesis and the one of thematic processes done by \cite{raimbault2018caracterisation} is the quasi absence here of models integrating governance processes. This may also be a direction to be explored as suggested by \cite{le2015modeling}.
	\item On the contrary, a very good correspondance can be established between geographical models of urban systems and the theoretical positioning of the evolutive urban theory introduced by \cite{pumain2018evolutionary}. This correspondance, more difficult to exhibit for all the other approaches reviewed, also suggests that this may be a relevant direction to follow.
\end{enumerate}

\section{Conclusion}

The processes linking transportation networks and territories are multi-scalar, hybrid and heterogenous. Therefore, the possible viewpoints and research questions are necessarily broad, complementary and rich. This could be a fundamental characteristic of socio-technical systems, which Pumain advocates for in~\cite{pumain2005cumulativite} as ``a new measure of complexity'', which would be linked to the number of viewpoints necessary to grasp a system at a given level of exhaustivity.

In order to better understand the neighbouring scientific landscape, and quantify the roles or relative weights of each, we have lead several bibliometrics and literature mapping analysis in~\cite{raimbault2018caracterisation}. A first preliminary analysis based on an algorithmic systematic review by \cite{raimbault2017models} suggests a certain compartmentalization of domains. This conclusion is confirmed by the citation and semantic network analysis of \cite{raimbault2018caracterisation}, which also allowed drawing disciplinary boundaries more precisely, both for the direct relations (citations) but also their scientific proximity for the terms and methods used. This paper completes these studies by using the constituted corpus and this knowledge of domains to achieve a semi-automatic systematic review, providing a corpus of papers directly dealing with interaction models. We then fully screened full texts, allowing to extract characteristics of models. A meta-analysis through statistical models linked these to the different domains. This provides a clear picture of what is done on this subject, why, how, and by which discipline. We finally discussed potential future directions for new interaction models, stemming from this analysis.








\section*{Supplementary material}

We give here the full numerical results of statistical analysis linking model characteristics and explicative variables.

\subsection*{Modalities of variables}

We recall here the variables used in the meta-analysis and their modalities. These are:

\begin{itemize}
	\item Type of model (\texttt{TYPE}): strong, territory, network.
	\item Publication year (\texttt{YEAR}), integer number.
	\item Citation community (\texttt{CITCOM}), defined within the citation network: Accessibility, Geography, Infra Planning, LUTI, Networks, TOD.
	\item A priori discipline (\texttt{DISCIPLINE}): biology, computer science, economics, engineering, environment, geography, physics, planning, transportation.
	\item Semantic community (\texttt{SEMCOM}): brt, complex networks, hedonic, hsr, infra planning, networks, tod.
	\item Methodology used: ca (Cellular Automaton), eq (analytical equations), map (cartography), mas (Multi-agent simulation), ro (operations research), sem (Structural Equation Modeling), sim (simulation), stat (statistics).
	\item Interdisciplinarity index (\texttt{INTERDISC}): real number in $[0,1]$.
	\item Temporal scale (\texttt{TEMPSCALE}): given in years, is set to 0 for static analyses.
	\item Spatial scale (\texttt{SPATSCALE}): continent (10000), country (1000), region (100), metro (10). These modalities are numerically transformed in km by the values given in parenthesis (stylized scales).
\end{itemize}

\subsection*{Model selection}

\begin{figure}
\includegraphics[width=\linewidth]{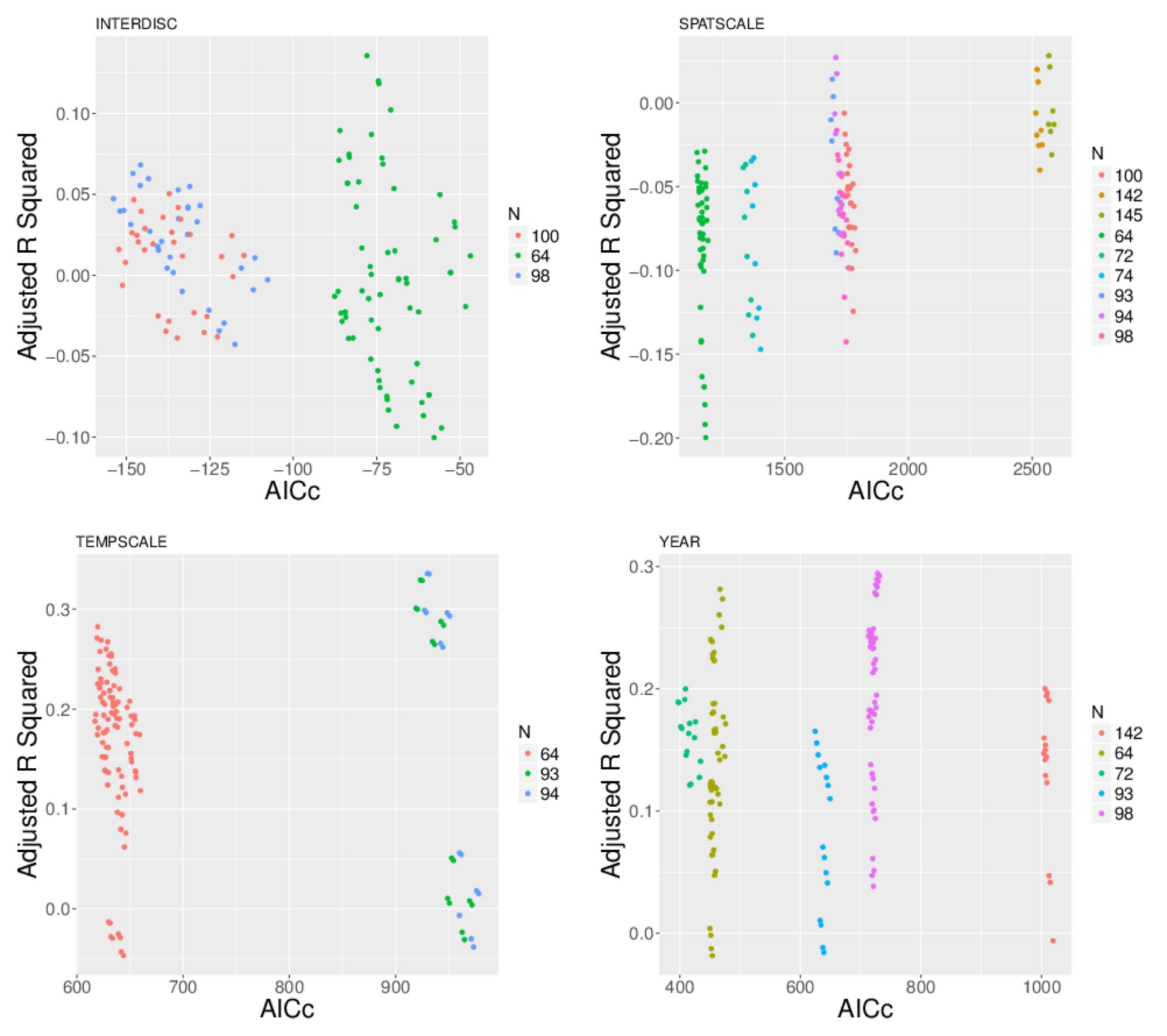}
\caption{\textbf{Multi-objective selection of linear models.} For each variable to be explained, we represent the position of all linear models in the objective space (corrected Akaike criterion AICc and adjusted $R^2$). The color of points gives the number of observations.\label{fig:app:quantepistemo:regressions}}
\end{figure}

Regarding model selection, it is not achieved following a unique criteria, because of the low number of observations for some models, but by the optimization in the Pareto sense of contradictory objectives of adjustment (adjusted $R^2$, to be maximized) and of the overfitting (corrected Akaike criterion AICc, to be minimized), while controlling the number of observation points. The Fig.~\ref{fig:app:quantepistemo:regressions} gives for each variable to be explained the localization of the set of potential models within the objective space, and also the corresponding number of observations. For interdisciplinarity, two point clouds correspond to different compromises, and we select the two optimal models (one for each cloud). For the spatial scale, we postulate a positive $R^2$, and a single optimal model then emerges. For the temporal scale, we have as for interdisciplinarity two compromise models. Finally for the year, the AICc gain between the two potential optima is negligible in comparison to the $R^2$ loss, and we thus select the optimal model such that $R^2>0.25$ and AICc$<600$. The results of models are given below.

\subsection*{Full results of statistical models}

\paragraph{Interdisciplinarity}

Interdisciplinarity is adjusted according to the linear models presented in Table~\ref{tab:app:modelography:interdisc}.

\begin{table}
\centering 
  \caption{\textbf{Linear models for interdisciplinarity}\label{tab:app:modelography:interdisc}}
\begin{tabular}{@{\extracolsep{5pt}}lcc} 
\footnotesize
\\[-1.8ex]\hline 
\hline \\[-1.8ex] 
\\[-1.8ex] & \multicolumn{2}{c}{INTERDISC} \\ 
\\[-1.8ex] & (1) & (2)\\ 
\hline \\[-1.8ex] 
 YEAR & $-$0.004 ($-$0.008, $-$0.00002), p = 0.055$^{*}$ & $-$0.002 ($-$0.005, 0.0001), p = 0.061$^{*}$ \\ 
  TEMPSCALE & $-$0.0003 ($-$0.001, 0.001), p = 0.615 &  \\ 
  DISCIPLINEengineering & 0.144 ($-$0.082, 0.371), p = 0.218 &  \\ 
  DISCIPLINEenvironment & 0.092 ($-$0.132, 0.316), p = 0.425 &  \\ 
  DISCIPLINEgeography & 0.036 ($-$0.043, 0.114), p = 0.378 &  \\ 
  DISCIPLINEphysics & $-$0.103 ($-$0.287, 0.080), p = 0.275 &  \\ 
  DISCIPLINEplanning & $-$0.047 ($-$0.135, 0.041), p = 0.30 &  \\ 
  DISCIPLINEtransportation & 0.062 ($-$0.025, 0.149), p = 0.169 &  \\ 
  TYPEstrong &  & $-$0.026 ($-$0.134, 0.081), p = 0.633 \\ 
  TYPEterritory &  & 0.044 ($-$0.026, 0.114), p = 0.222 \\ 
  SEMCOMcomplex networks &  & $-$0.217 ($-$0.522, 0.087), p = 0.166 \\ 
  SEMCOMhedonic & $-$0.179 ($-$0.407, 0.049), p = 0.130 & $-$0.184 ($-$0.400, 0.032), p = 0.100$^{*}$ \\ 
  SEMCOMhsr & $-$0.100 ($-$0.361, 0.162), p = 0.459 & $-$0.122 ($-$0.357, 0.112), p = 0.309 \\
  SEMCOMinfra planning & $-$0.032 ($-$0.273, 0.209), p = 0.797 & $-$0.096 ($-$0.321, 0.128), p = 0.404  \\ 
  SEMCOMnetworks & $-$0.038 ($-$0.272, 0.195), p = 0.750 & $-$0.107 ($-$0.324, 0.109), p = 0.335 \\ 
  SEMCOMtod & $-$0.105 ($-$0.332, 0.121), p = 0.366 & $-$0.152 ($-$0.364, 0.060), p = 0.165 \\ 
  Constant & 8.962 (0.776, 17.147), p = 0.037$^{**}$ & 5.531 (0.575, 10.487), p = 0.032$^{**}$ \\
 \hline \\[-1.8ex] 
Observations & 64 & 98 \\ 
R$^{2}$ & 0.314 & 0.155 \\ 
Adjusted R$^{2}$ & 0.136 & 0.068 \\ 
Residual Std. Error & 0.109 (df = 50) & 0.107 (df = 88) \\ 
F Statistic & 1.761$^{*}$ (df = 13; 50) & 1.789$^{*}$ (df = 9; 88) \\ 
\hline 
\hline \\[-1.8ex] 
\textit{Note:}  & \multicolumn{2}{r}{$^{*}$p$<$0.1; $^{**}$p$<$0.05; $^{***}$p$<$0.01} \\ 
\end{tabular}
\end{table} 

\paragraph{Spatial scale}

The spatial scale is adjusted following the linear model which adjustment is given in Table~\ref{tab:app:modelography:spatscale}.

\begin{table}
\centering 
  \caption{\textbf{Linear model for the spatial scale.}\label{tab:app:modelography:spatscale}}
\begin{tabular}{@{\extracolsep{5pt}}lc} 
\\[-1.8ex]\hline 
\hline \\[-1.8ex] 
\\[-1.8ex] & SPATSCALE \\ 
\hline \\[-1.8ex] 
 TEMPSCALE & $-$5.179 ($-$16.259, 5.901) \\ 
  & p = 0.363 \\ 
  DISCIPLINEengineering & $-$154.461 ($-$3,003.326, 2,694.405) \\ 
  & p = 0.916 \\ 
  DISCIPLINEenvironment & $-$5.878 ($-$3,977.974, 3,966.219) \\ 
  & p = 0.998 \\ 
  DISCIPLINEgeography & 1,445.457 (389.349, 2,501.565) \\ 
  & p = 0.009$^{***}$ \\ 
  DISCIPLINEphysics & 292.559 ($-$2,717.659, 3,302.777) \\ 
  & p = 0.850 \\ 
  DISCIPLINEplanning & $-$143.554 ($-$1,361.357, 1,074.249) \\ 
  & p = 0.818 \\ 
  DISCIPLINEtransportation & 568.329 ($-$606.167, 1,742.826) \\ 
  & p = 0.346 \\ 
  Constant & 235.357 ($-$458.201, 928.914) \\ 
  & p = 0.508 \\ 
 \hline \\[-1.8ex]
Observations & 94 \\ 
R$^{2}$ & 0.100 \\ 
Adjusted R$^{2}$ & 0.027 \\ 
Residual Std. Error & 1,995.272 (df = 86) \\ 
F Statistic &  1.369 (df = 7; 86) \\ 
\hline 
\hline \\[-1.8ex] 
\textit{Note:}  & \multicolumn{1}{r}{$^{*}$p$<$0.1; $^{**}$p$<$0.05; $^{***}$p$<$0.01} \\ 
\end{tabular}
\end{table} 

\paragraph{Time scale}

The temporal scale is adjusted according to the linear models presented in Table~\ref{tab:app:modelography:tempscale}.

\begin{table}
\centering 
    \caption{\textbf{Linear models for the temporal scale.}\label{tab:app:modelography:tempscale}}
\begin{tabular}{@{\extracolsep{5pt}}lcc} 
\\[-1.8ex]\hline 
\hline \\[-1.8ex] 
\\[-1.8ex] & \multicolumn{2}{c}{TEMPSCALE} \\ 
\\[-1.8ex] & (1) & (2)\\ 
\hline \\[-1.8ex] 
 YEAR & 0.674 ($-$0.294, 1.643) &  \\ 
  & p = 0.179 &  \\ 
  TYPEstrong &  & 100.271 (58.312, 142.230) \\ 
  &  & p = 0.00002$^{***}$ \\ 
  TYPEterritory & $-$38.933 ($-$64.249, $-$13.617) & $-$14.988 ($-$37.411, 7.435) \\ 
  & p = 0.004$^{***}$ & p = 0.194 \\ 
  DISCIPLINEengineering & $-$52.107 ($-$110.950, 6.735) & $-$9.609 ($-$55.841, 36.624) \\ 
  & p = 0.089$^{*}$ & p = 0.685 \\ 
  DISCIPLINEenvironment & 17.110 ($-$37.350, 71.569) & 17.886 ($-$45.319, 81.090) \\ 
  & p = 0.541 & p = 0.581 \\ 
  DISCIPLINEgeography & 3.640 ($-$15.364, 22.644) & 9.126 ($-$7.590, 25.843) \\ 
  & p = 0.709 & p = 0.288 \\ 
  DISCIPLINEphysics & 46.879 (0.638, 93.120) & 77.897 (28.225, 127.570) \\ 
  & p = 0.053$^{*}$ & p = 0.003$^{***}$ \\ 
  DISCIPLINEplanning & 1.304 ($-$19.336, 21.945) & 4.553 ($-$14.865, 23.971) \\ 
  & p = 0.902 & p = 0.648 \\ 
  DISCIPLINEtransportation & $-$14.718 ($-$34.978, 5.543) & 8.753 ($-$9.864, 27.371) \\ 
  & p = 0.161 & p = 0.360 \\ 
  INTERDISC & 2.357 ($-$59.200, 63.915) &  \\ 
  & p = 0.941 &  \\ 
  Constant & $-$1,305.126 ($-$3,252.499, 642.247) & 22.103 ($-$0.951, 45.156) \\ 
  & p = 0.195 & p = 0.064$^{*}$ \\ 
 \hline \\[-1.8ex] 
Observations & 64 & 94 \\ 
R$^{2}$ & 0.385 & 0.393 \\ 
Adjusted R$^{2}$ & 0.282 & 0.336 \\ 
Residual Std. Error & 26.984 (df = 54) & 31.747 (df = 85) \\ 
F Statistic & 3.755$^{***}$ (df = 9; 54) & 6.871$^{***}$ (df = 8; 85) \\ 
\hline 
\hline \\[-1.8ex] 
\textit{Note:}  & \multicolumn{2}{r}{$^{*}$p$<$0.1; $^{**}$p$<$0.05; $^{***}$p$<$0.01} \\ 
\end{tabular} 
\end{table} 

\paragraph{Year}

The publications year is adjusted following the linear model which adjustement is given in Table~\ref{tab:app:modelography:year}.

\begin{table}
\centering
  \caption{\textbf{Linear model for the publication year.}\label{tab:app:modelography:year}}
\begin{tabular}{@{\extracolsep{5pt}}lc} 
\footnotesize
\\[-1.8ex]\hline 
\hline \\[-1.8ex] 
\\[-1.8ex] & YEAR \\ 
\hline \\[-1.8ex] 
 TYPEterritory & 10.898 (3.045, 18.750), p = 0.010$^{***}$ \\ 
  TEMPSCALE & 0.035 ($-$0.033, 0.103), p = 0.320 \\ 
  FMETHODeq & $-$6.224 ($-$20.162, 7.714), p = 0.387 \\ 
  FMETHODmap & 4.747 ($-$7.595, 17.089), p = 0.456 \\ 
  FMETHODro & 6.128 ($-$11.694, 23.950), p = 0.504 \\ 
  FMETHODsem & 1.009 ($-$16.659, 18.676), p = 0.912 \\ 
  FMETHODsim & 5.153 ($-$6.809, 17.114), p = 0.404 \\ 
  FMETHODstat & $-$0.357 ($-$10.925, 10.211), p = 0.948 \\ 
  DISCIPLINEengineering & 13.486 ($-$7.238, 34.210), p = 0.210 \\ 
  DISCIPLINEenvironment & $-$3.668 ($-$21.605, 14.269), p = 0.691 \\ 
  DISCIPLINEgeography & 1.121 ($-$4.528, 6.769), p = 0.700 \\ 
  DISCIPLINEphysics & 3.392 ($-$8.461, 15.245), p = 0.578 \\ 
  DISCIPLINEplanning & $-$2.850 ($-$8.873, 3.173), p = 0.359 \\ 
  DISCIPLINEtransportation & 5.503 (0.006, 11.000), p = 0.057$^{*}$ \\ 
  INTERDISC & $-$12.876 ($-$29.567, 3.815), p = 0.138 \\ 
  SEMCOMhedonic & $-$5.769 ($-$19.931, 8.393), p = 0.430 \\ 
  SEMCOMhsr & 6.135 ($-$9.889, 22.159), p = 0.458 \\ 
  SEMCOMinfra planning & $-$4.123 ($-$18.910, 10.663), p = 0.588 \\ 
  SEMCOMnetworks & 4.711 ($-$9.736, 19.158), p = 0.527 \\ 
  SEMCOMtod & $-$1.653 ($-$15.837, 12.532), p = 0.821 \\ 
  Constant & 2,004.945 (1,981.531, 2,028.359), p = 0.000$^{***}$ \\ 
 \hline \\[-1.8ex]
Observations & 64 \\ 
R$^{2}$ & 0.510 \\ 
Adjusted R$^{2}$ & 0.281 \\ 
Residual Std. Error & 6.617 (df = 43) \\ 
F Statistic & 2.234$^{**}$ (df = 20; 43) \\ 
\hline 
\hline \\[-1.8ex] 
\textit{Note:}  & \multicolumn{1}{r}{$^{*}$p$<$0.1; $^{**}$p$<$0.05; $^{***}$p$<$0.01} \\ 
\end{tabular}
\end{table} 

\end{document}